\documentclass[twocolumn]{aa}

\usepackage{graphicx}
\usepackage{natbib}
\usepackage{txfonts}
\bibpunct{(}{)}{;}{a}{}{,}

\newcommand{\fei}[1]{{Fe\,{\sc i}\,$\lambda$}#1}
\newcommand{\feis}[1]{{Fe\,{\sc i}}#1}

\begin{document}
\title{Magnetic field distribution in the quiet Sun:\\
		a simplified model approach}

\author{F.~Berrilli \and D.~Del Moro \and B.~Viticchi\`{e}}

\institute{Dipartimento di Fisica, Universit\`{a} degli Studi di Roma ``Tor Vergata'',
	Via della Ricerca Scientifica 1, I-00133 Roma, Italy\\
	\email{viticchie@roma2.infn.it}}

\abstract
{The quiet Sun presents magnetized plasma whose field strengths vary from zero to about $2$~kG. The probability density function of the magnetic field strength $B$ efficaciously describes the statistical properties of the quiet Sun magnetic field.}
{We simulate the dynamics and the evolution of quiet Sun magnetic elements to produce a probability density function of the field strengths associated with such elements.}
{The dynamics of the magnetic field are simulated by means of a numerical model in which magnetic elements are driven passively by an advection field characterized by spatio-temporal correlations that mimick the granulation and mesogranulation scales observed on the solar surface. The field strength can increase due to an amplification process that occurs where the magnetic elements converge. Starting from a $\delta$-like probability density function centered on $B=30$~G, we obtain magnetic field strengths of up to $2$~kG (in absolute value). To derive the statistical properties of the magnetic elements, several simulation runs are performed.}
{Our model is able to produce kG magnetic fields in a time interval of the order of the granulation timescale. The mean unsigned flux density and the mean magnetic energy density of the synthetic quiet Sun reach values of $\langle B \rangle \simeq 100$~G and $\langle B^2 \rangle^{1/2}\simeq 350$~G respectively in the stationary regime. The derived probability density function of the magnetic field strength decreases rapidly from $B=30$~G to $B \sim 100$~G and has a secondary maximum at $B=2$~kG. From this result, it follows that magnetic fields $\ge700$~G dominate the unsigned flux density and magnetic energy density, although the probability density function of the field strength reaches a maximum at $B \sim 10$~G.}
{A photospheric advection field with spatio-temporal correlations, driving the magnetic elements, and reduced magnetic amplification rules are able to create a realistic probability density function of the quiet Sun magnetic field. It has been found that they naturally produce an excess of magnetic fields around $2$~kG if an upper limit is imposed on the field strength.}

\titlerunning{Quiet Sun PDF: a simplified model approach}
\authorrunning{F.~Berrilli, D.~Del Moro \and B.~Viticchi\'{e}}
\maketitle

\section{Introduction}
\label{intro}
When observing the Sun in full disk magnetograms, magnetic concentrations define the network pattern
associated with the supergranular motions of photospheric plasma, while the interior of the network pattern 
appears to be unmagnetized.\\
We know that weak polarization signals are measurable almost everywhere in the interior of network cells  \citep{livhar75,smi75}. These signals are associated with the internetwork component of the solar magnetic field, also called the quiet Sun magnetic field, which covers more than 90\% of the solar surface, independent of the solar activity \citep{harang93}.
Due to its large diffusion onto the solar surface, the quiet Sun magnetic field may be useful for understanding the physics of the photospheric magnetism. In fact, it could hold a large fraction of the unsigned flux density and magnetic
energy density of the solar photosphere \citep[e.g.][]{unn59,ste82,san98,san04,schtit03}.\\
The weak polarization signals measured in quiet Sun regions can be interpreted as the result of the linear 
combination of the polarization emerging from discrete magnetic flux tubes, which are typically smaller than the angular
resolution, that fill the solar atmosphere with a complex topology \citep[e.g.][]{emocat01a,san03}.\\
Knowledge of the quiet Sun has been enhanced significantly by improvements in instrumentation \citep[e.g.][]{san96,linrim99,lit02,dom03a,dom03b,kho03}, in diagnostic techniques \citep[e.g.][]{ste82,fausch93,fausch95,lan98,ree00,sanlit00,soc01,socsan02,san04,tru04,man04,san05,mar06b}, and numerical simulations \citep[e.g.][]{cat99,emocat01b,stenor02,vog03th,vog05,
stenor06,vogsch07}.\\
These improvements were crucial in characterizing the quiet Sun magnetic field and its different observational
parameters. One parameter, the fraction of solar surface occupied by a certain field strength, is a very powerful tool. In fact, this can be trivially translated into the probability density function (PDF) of the magnetic field strength, $P(B)$, which is useful
in describing the statistical properties of a quiet Sun region. For example, calculating the first and the second
momentum of $P(B)$ provides information about the mean unsigned flux density and the mean magnetic energy density (apart from a
constant factor $1/8\pi$), respectively. Using a PDF assumes implicitly a continuum for values of the magnetic field strength. The limit imposed by the pressure equilibrium between the magnetic flux tubes and photospheric plasma also defines the variability range for quiet Sun magnetic fields to be between $0$~G to about $2$~kG.\\
\citet{lin95} computed the PDF in active and quiet Sun regions using \fei{15648} and \fei{15652} lines, while \citet{lit02} obtained it from the analysis of network regions and their surroundings using \feis{} lines at $6300$~\AA{}. Both authors found a bimodal distribution with a dominant contribution at $\simeq 150$~G and a secondary peak at $\simeq 1.5$~kG. In contrast, \citet{col01}, studying quiet Sun regions using the same infrared lines as \citet{lin95}, found an exponential decline in the magnetic flux density histogram from a peak at $\simeq 200$~G and no counts for magnetic flux density $\ga 1$~kG. \citet{dom06} developed a set of PDFs for the quiet Sun magnetic fields that was consistent with observations and magnetoconvection simulations by using information from both Hanle and Zeeman observations; their study created substantial interest because the functions increased for strong fields, just before a cut-off at $B\simeq 1.8$~kG. This characteristic implies that strong fields are important because, even when present in a small fraction of the photosphere, they dominate the unsigned flux density and magnetic energy density. On the other hand, the magnetic field distributions proposed by \citet{mar06a,mar06b} and \citet{mar07}, which were derived from simultaneous infrared and visible observations of \feis{} lines, do not present any rise for strong fields. The inversion of the \fei{6301} and \fei{6302} profiles observed by the \textit{HINODE} satellite \citep{oro07} confirms this result.\\
Magnetoconvection simulations provide a wide variety of quiet Sun magnetic field PDFs to be compared with those derived from observations. Simulations show, for example, exponential decays \citep{cat99,ste03,stenor06,vogsch07}, bimodal behaviors \citep{vog03,vogsch03}, or an ``exponential-like core and a modest peak'' for $B \sim 1$~kG \citep{ser03}.\\
An important step is to ascertain the presence or absence of a bump for magnetic fields $\sim 1$~kG in quiet Sun magnetic field strength PDFs; since the magnetic flux and magnetic energy both scale as powers of the field strength, kG magnetic fields should play a crucial role in defining the global properties of quiet Sun magnetic fields. The connectivity between the lower and the upper solar atmosphere may indeed be defined by the larger field strength concentrations in the photosphere \citep{dom06}.\\
\citet{san07} indicated that a bump in the PDF for quiet Sun fields $\sim$~kG is to be expected if ``a magnetic amplification mechanism operates in the quiet Sun''. The author derived an equation for the temporal evolution of the PDF, whose shape can be modified by magnetic amplification and photospheric convection. In the stationary state, the model is characterized by a $P(B)$ with a bump for strong fields.\\
In our model, we implement the dynamics of a high number of photospheric magnetic elements, driven by a photospheric velocity field in which a large-scale organization (mesogranulation) is naturally derived from the cooperation of small-scale flows (granulation). The model also takes into account simplified processes that are able to amplify and limit the magnetic field strength. In \S~\ref{model}, we describe the model, in \S~\ref{pdfs}, the results of the simulation are presented, while in \S~\ref{disc}, they are discussed. Finally, in \S~\ref{conc}, the conclusions are outlined.
\section{The model}
\label{model}
\begin{figure}
\centering
  \includegraphics[width=8cm]{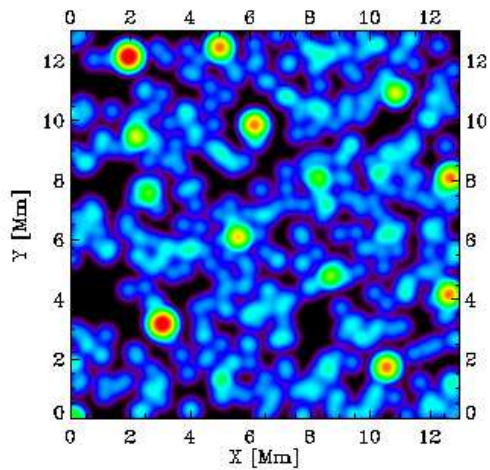}
    \caption{Illustrative representation of $\vert B \vert$ in the computational domain. Black areas represent $0$~G regions; blue and sky-blue  represent $\sim 10$~G regions; green represents $\sim 100$~G regions; yellow and red represent $\sim$~kG magnetic concentrations. The position of concentrations with $B\sim$~kG correspond to the position of stable downward plumes organized over mesogranular scales.}
    \label{modelpic}
\end{figure}
The model simulates the dynamics and evolution of photospheric magnetic elements driven by a photospheric velocity field characterized by spatio-temporal correlations. To reproduce these correlations, we simulated the photospheric velocity field in a way similar to that proposed by \citet{ras03}; it was computed by an \textit{n}-body simulation that realizes the interaction between downflow plumes by considering only their mutual horizontal interaction, as in \citet{ras98}. The movement of each plume was determined by the advection produced by the other plumes. This model spontaneously creates stable downward plumes, organized on a mesogranular pattern, from the interaction of granular scale flows. The ratios between the mesogranular and granular spatio-temporal scales, over which the model self-organizes, agree with the ratios measured in the solar photosphere. \citet{vit06} utilized such a velocity field to drive magnetic loop footpoints in a reconnection model that reproduces the observed statistical properties nanoflares.\\
\indent In the present work, the simulations start with the advection field only, without any magnetic element in the domain. When the velocity field has reached a stationary behavior, i.e. after the onset of mesogranular scale flows, the time $t$ is set to zero and $N_{0}=500$ magnetic elements are distributed at random positions over the computational domain. The magnetic elements have an associated field strength $B_{in}=30$~G with random orientation ($\pm1$) and size equal to the spatial resolution element. Hereafter, $B$ represents the absolute value of the magnetic field strength. We choose such a value for $B_{in}$ since it is approximately equal to the minimum magnetic flux density measurable nowadays \citep[e.g.][]{tru04,bom05,car08}.
We allow the magnetic elements to be transported passively by the photospheric velocity field towards different plume sites. At each plume site, a certain number of magnetic elements converge under the action of the advection field. When two magnetic elements, associated with $B_1$ and $B_2$, overlap in the same resolution element, they can produce the following two results:\\
\indent 1. If the two elements have opposite orientations, they are substituted by a new magnetic element at the same site with an associated $B=\vert B_1-B_2\vert$ and the orientation of the strongest element. The total flux is conserved such that the spatial size of the new element remains as one resolution element. In the particular case of $B_1=B_2$, the elements cancel the effect of each other.\\
\indent 2. If the two elements have the same orientation, they are substituted by a new magnetic element in the same site with an associated $B=B_1+B_2$ and the same orientation. The total flux is again conserved and the spatial size of the new element is one resolution element.\\ 
This process implicates an amplification of the magnetic field strength. Different mechanisms have been proposed to concentrate magnetic fields in quiet Sun regions up to the kG range \citep[e.g.][]{wei66,par78,spr79,san01}. As reported by \citet{san07}, all mechanisms are abruptly diminished at the maximum field strength imposed by the gas pressure of the quiet Sun photosphere. Therefore, we place the constraint that the amplification process cannot create field strengths greater than $B_{lim}=2$~kG. We can interpret $B_{lim}$ as a representative value of the maximum field strengths for quiet Sun regions \citep[e.g.][]{par78,spr79,grd98,ste03,dom06,san07}. Whenever the amplification process would produce $B>B_{lim}$, the magnetic elements do not interact and remain separated.\\
We consider implicitly that the amplification mechanism operates on the same timescales as the granulation timescale ($\sim10$~min). This hypothesis is consistent with the amplification timescales reported by \citet{par78} and \citet{grd98}. Up to the equipartition value of about $500$~G, the amplification can be ascribed to the granular flows and five minutes are sufficient to produce kG fields from a uniform distribution of $400$~G.\\
As proposed by \citet{ser03}, starting from an initial PDF of $P_{t=0}(B)=\delta(B-B_{in})$, the amplification process will broaden the PDF and extend it to $B_{lim}$.\\
To simulate the emergence of new magnetic elements on the solar surface, we continuously add randomly located elements of random orientation and $B_{in}$ magnetic field strength. This injection process maintains the total number of magnetic elements $\sim N_0$ and compensates for the reduction in the total number due to the concentration process.\\
Figure~\ref{modelpic} shows an illustrative representation of a snapshot of the simulation in the stationary regime: we represent the $\vert B \vert$ pattern emerging on the computational domain.
\section{PDF of the quiet Sun magnetic field strengths}
\label{pdfs}
\begin{figure}
  \resizebox{\hsize}{!}{\includegraphics{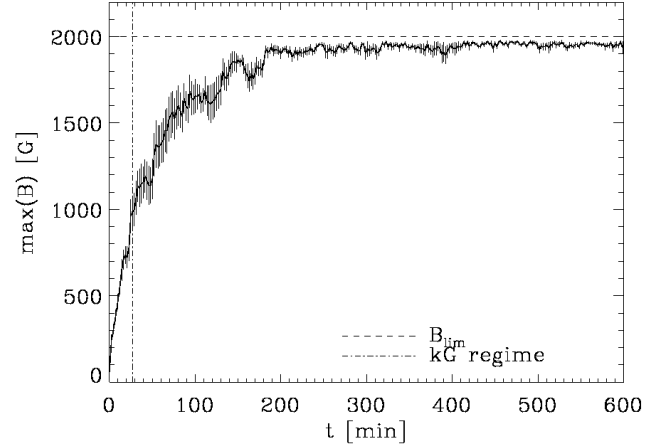}}
  \caption{Average maximum field strength versus evolution time $t$. The function is obtained by averaging the results from $n_{RUN}=10$ distinct simulation runs and the error bars are calculated to equal $\frac{\sigma}{\sqrt{n_{RUN}}}$. The horizontal dashed line represents the maximum value allowed for the magnetic flux density ($B_{lim}$), while the vertical dot-dashed line indicates the instant at which the model produces the first kG element. The plot shows only a part of the total time domain ($t\simeq2600$~min).}
  \label{meanmax}
\end{figure}
\begin{figure}
  \resizebox{\hsize}{!}{\includegraphics{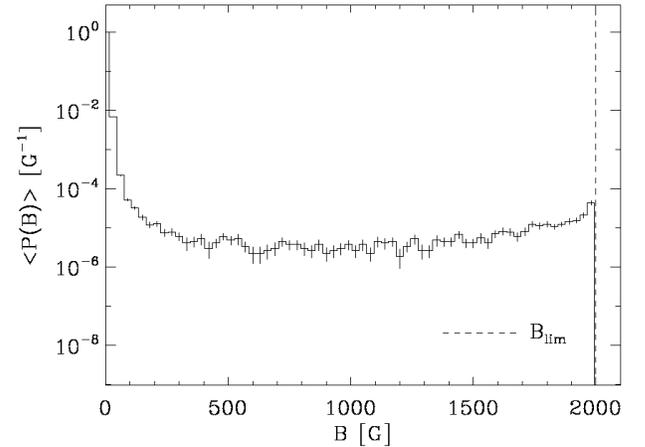}}
  \caption{Mean probability density function $\langle P(B)\rangle$ for the magnetic field strength in the stationary regime. $\langle P(B)\rangle$ has been obtained averaging $n_{TOT}=40$ non-correlated PDFs and the error bars are calculated to equal $\frac{\sigma}{\sqrt{n_{TOT}}}$. The histogram bin-size is $30$~G. The dashed line represents the upper limit for the magnetic amplification process $B_{lim}$. The maximum at $B=0$~G is produced by the fraction of the domain free of magnetic elements.}
  \label{meanpdf}
\end{figure}
\begin{figure}
  \resizebox{\hsize}{!}{\includegraphics{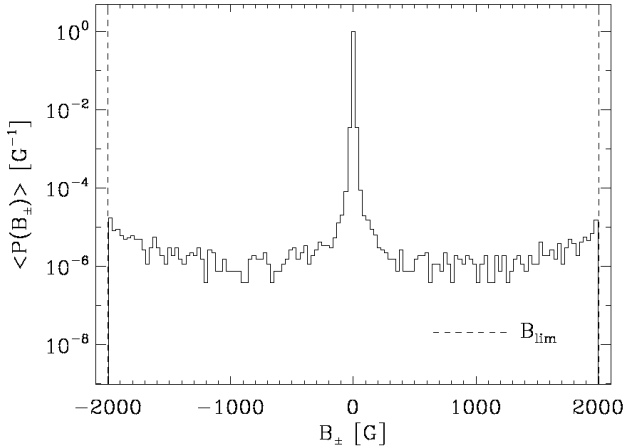}}
  \caption{Mean probability density function $\langle P(B_\pm)\rangle$ for the signed magnetic field strength in the stationary regime. The histogram bin-size is $30$~G. The vertical dashed lines represent the upper and lower limits of the magnetic amplification process $\pm B_{lim}$. The maximum at $B=0$~G is produced by the fraction of the domain free of magnetic elements.}
  \label{signedpdf}
\end{figure}
To derive a PDF of the quiet Sun magnetic field strengths, we performed ten simulation runs with different advection fields and different initial distributions of magnetic elements. Each simulation reproduced the evolution of the quiet Sun for an equivalent time of about $2600$~min. We considered the behavior of the model by analyzing its average maximum field strength $max(B)$, mean unsigned flux density $\langle B \rangle$, and mean magnetic energy density $\langle B^2 \rangle/8\pi$, as functions of the mean evolution time $t$:
\begin{eqnarray*}
<B>=\int_{0}^{B_{lim}}BP(B)dB,~\\
<B^2>=\int_{0}^{B_{lim}}B^2P(B)dB,
\end{eqnarray*}
in a similar way to \citet{dom06}.\\ 
The model creates magnetic concentrations with $B=1$~kG in about $30$~min and $B\simeq B_{lim}$ in about $200$~min (Fig.~\ref{meanmax}).\\
The quantity $max(B)$ is unsuitable for describing the stationary regime of the magnetic element system because its value is limited. As discussed in \S~\ref{model}, we also expect to obtain a PDF spanning the entire domain $B=~0-B_{lim}$; the stationary regime must therefore be identified referring to the integral properties of the PDF, such as the mean unsigned flux density and the mean magnetic energy density. $\langle B \rangle$ and $\langle B^2 \rangle^{1/2}$ saturate exponentially at $\langle B \rangle_{sat}\simeq 100$~G and $\langle B^2 \rangle_{sat}^{1/2}\simeq 350$~G, respectively. We interpret this saturation as an indication that the entire system has reached the stationary regime. In this regime, we compute $\langle P(B)\rangle$ by averaging forty non-correlated PDFs. The computed $\langle P(B)\rangle$ is shown in Fig.~\ref{meanpdf}.\\
Alternatively, we could compute the mean probability density function $\langle P(B_\pm)\rangle$ for the signed magnetic field strength by considering the direction of the magnetic fields (shown in Fig.~\ref{signedpdf}). As expected, the $\langle P(B_\pm)\rangle$ is quite symmetric with respect to zero.
\section{Discussion}
\label{disc}
\begin{figure}[t!]
\centering
  \includegraphics[width=9.2cm]{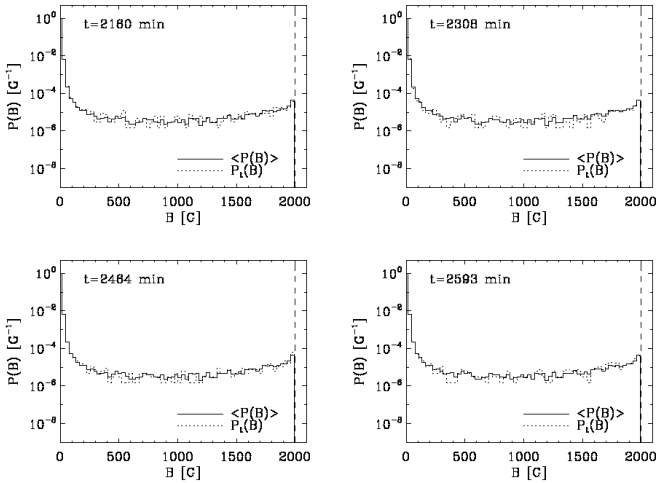}
  \caption{Mean probability density function $\langle P(B)\rangle$ (solid line, same as Fig.\ref{meanpdf}) compared with four $P_t(B)$ (dotted lines) calculated at different instants of the simulation. Each $P_t(B)$ represents the mean of ten PDFs from ten simulation runs at time $t$. \textit{Top left}:$t=2160$~min, \textit{top right} $t=2308$~min, \textit{bottom left} $t=2464$~min and \textit{bottom right} $t=2593$~min. The histogram bin-size is $30$~G. The vertical dashed lines represent the upper and lower limits of the magnetic amplification process $\pm B_{lim}$. The maximum at $B=0$~G is produced by the fraction of the domain free of magnetic elements.}
  \label{difference}
\end{figure}

In the previous section we presented the results of several simulations of our reduced model. We now focus on some particular points in detail.\\
The maximum field strength, plotted in Fig.~\ref{meanmax}, as a function of $t$ indicates that the timescale over which kG fields are produced is approximately $30$~min. This is the same timescale over which the mesogranular pattern emerged in the numerical simulation of \citet{vit06}.\\
Different descriptions of the magnetic amplification process \citep{grd98,san01} achieved similar results: weak magnetic-field structures, in a downflow environment, develop field strengths $\sim$~kG within a few minutes. \citet{ber05} reported that the organization of structures on mesogranular boundaries is rapid with a typical timescale of $\simeq10$~min.\\
The analysis of the position of strong field magnetic elements in our simulation revealed that these are superimposed onto stable downward plumes on mesogranular scales (Fig.~\ref{modelpic}). Such a correlation between strong magnetic fields and the mesogranular advection scale is in agreement with \citet{dom03}, who found a correlation between mesogranular photospheric flows and the pattern observed in magnetograms.\\
The symmetry of the $\langle P(B_\pm)\rangle$, reported in Fig.~\ref{signedpdf}, confirms the equal probability of having concentrations of positive and negative direction. This originates in both the initial conditions and the new element addition criteria applied during the evolution. Similar symmetries have been found in simulations \citep[e.g.,][]{vogsch07} and in observations of \citet{mar06a,mar07}. In the latter works the authors pointed out that such a symmetry may imply a local dynamo origin for the internetwork magnetic field, that is a magnetic field controlled by solar granulation.\\
The sharp count decay found at $B=B_{lim}$ is obviously due to the ``interaction turn off'' produced by $B_{lim}$ (\S~\ref{model}). Other PDFs reported in the literature show a smooth decay of the peak for $B \simeq 1.8$~kG \citep{dom06,san07}, as expected for natural PDFs. We could reproduce this smooth decay of $\langle P(B)\rangle$ for $B\gtrsim B_{lim}$ by modifying the magnetic element interaction probability, but it is not within the scope of this work to reproduce these fine details of the PDF.\\
To verify that the stationary regime has been reached, we compared the $\langle P(B) \rangle$ with four mean PDFs, $P_t(B)$, representative of four different instants of the simulation during the stationary regime (Fig.~\ref{difference}). Each $P_t(B)$ represents the mean of ten PDFs at a given time $t$. We consider the four $P_t(B)$ to be uncorrelated because they are separated by more than $140$ minutes, a time interval that is longer than three times the mean mesogranular lifetime. As shown in Fig.~\ref{difference}, the deviation between the $\langle P(B) \rangle$ and the four $P_t(B)$ shows no time dependence and the root mean square deviations of all four $P_t(B)$ with respect to the $\langle P(B) \rangle$ are approximately one part in a thousand.\\
The values $\langle B \rangle_{sat} \simeq 100$~G and $\langle B^2 \rangle_{sat}^{1/2}\simeq 350$~G, reached in the stationary regime, are in good agreement with the results of \citet{dom06}. The percentages of unsigned flux density and magnetic energy density for $B \ge 700$~G are $\simeq 65\%$ and $\simeq 97\%$ respectively, indicating a dominant role for kG fields. The percentages that we found are higher than those reported by \citet{dom06}; these differences can be explained by referring to the steeper decrease in the $\langle P(B)\rangle$ for the $0-700$~G interval with respect to the log-normal trend found by these authors.\\
The $\langle P(B)\rangle$ presents a maximum at $B=B_{in}$ in agreement with \citet{dom06}. This result originates in the injection value for the magnetic field strength $B_{in}=30$~G. At $B\simeq B_{lim}$, the PDF presents a secondary maximum, as found by \citet{ste03}, \citet{dom06}, and \citet{san07}.\\
In a similar way to \citet{san07}, we may conclude that this secondary maximum in the $\langle P(B)\rangle$ is due to the upper limit placed on the amplification process at $B_{lim}$. When a magnetic amplification process operates and a limit is placed on this process, an accumulation of field strength about the limit is expected. All assumptions of the model contribute in shaping the $\langle P(B) \rangle$: the strong convergence regions of the granular advection pattern create suitable conditions for amplifying the magnetic field and, at the same time, support the amplification process by gathering new magnetic elements to counteract the cancellation process. The secondary maximum of the PDF is produced by both the spatial and temporal correlations of the velocity field, which advect the magnetic elements, and the upper limit of the amplification process at $B_{lim}$.
\section{Conclusions}
\label{conc}
We have presented a model that reproduces the dynamics and evolution of photospheric magnetic elements. These elements are passively driven by an horizontal velocity field characterized by spatio-temporal correlations that agree with those observed in the solar photosphere. The description of the velocity field follows \citet{ras98,ras03}.
In their evolution, the magnetic elements interact and increase their magnetic field strengths through an amplification process assumed to work on a timescale comparable with the granular timescale and able to recreate kG concentrations. The injection scale, constrained to be $B_{in}=30$~G, and the upper limit at $B_{lim}=2$~kG are both derived from observations and physical constraints \citep[e.g.][]{tru04,par78}.\\
We investigated the statistical properties of the field strengths associated with the magnetic elements using a series of numerical simulations. The main conclusions are: 
\begin{enumerate}
\item The model produces kG fields in a time interval of the order of the mesogranulation timescale. These strong fields are organized in a pattern superimposed on the mesogranulation pattern of the advection field.
\item In the stationary regime, the PDF of the magnetic field strength has a stable shape and has a secondary maximum that corresponds to the upper limit set at $2$~kG.
\item The associated mean unsigned flux density and mean magnetic energy density in the stationary regime are $\langle B \rangle_{sat} \simeq 100$~G and $\langle B^2 \rangle_{sat}^{1/2}\simeq350$~G, respectively.
\item The magnetic element system described by this PDF is strongly dominated by kG magnetic fields: the fractions of unsigned flux density and magnetic energy density for $B\ge700$~G are $\simeq 65\%$ and $\simeq 97\%$, respectively.
\end{enumerate}
The quiet Sun is certainly more complex than the model presented here. However, a more simplified approach, based on the advection field proposed by \citet{ras03}, reproduced the observed distributions for the released magnetic energy and waiting times of nanoflare events \citep{vit06}. In the present work, we showed how the same approach can also be applied to reproduce some of the statistical properties of the quiet Sun magnetic field.

\begin{acknowledgements}
We are very grateful to the anonymous referee for invaluable comments on the manuscript. We thank Jorge S\'anchez Almeida for fruitful discussions about the topic. This paper was partially supported by Regione Lazio CVS (Centro per lo studio della variabilità\`a del Sole) PhD grant and ASI/INAF I/015/07/0 ESS grant.
\end{acknowledgements}

\bibliography{fullbib}
\bibliographystyle{aa} 

\end{document}